\documentclass[preprint,preprintnumbers,amsmath,amssymb]{revtex4-2}
\usepackage{amsfonts}    
\usepackage{amssymb}
\usepackage{latexsym}
\usepackage{graphicx}
\usepackage{rotating}
\usepackage{multirow}
\usepackage[usenames,dvipsnames]{color}
\usepackage[active]{srcltx}

\usepackage{hyperref}
\usepackage{color}

\newcommand{\al}{\alpha}
\newcommand{\pa}{\partial}

\newcommand{\veps}{\varepsilon}

\newcommand{\la}{\lambda}

\newcommand{\rar}{\rightarrow}

\begin{document}

\title{On the connection between perturbation theory and new semiclassical expansion in quantum mechanics}

\date{\today}

\author{A.V.~Turbiner$^{1,2}$}
\email{turbiner@nucleares.unam.mx, alexander.turbiner@stonybrook.edu}

\author{E.~Shuryak$^2$}
\email{edward.shuryak@stonybrook.edu}

\affiliation{$^1$ Instituto de Ciencias Nucleares, Universidad Nacional Aut\'onoma de M\'exico,
Apartado Postal 70-543, 04510 M\'exico, D.F., M\'exico}

\affiliation{$^2$  Department of Physics and Astronomy, Stony Brook University,
Stony Brook, NY 11794-3800, USA}

\begin{abstract}
It is shown that for the one-dimensional anharmonic oscillator with potential
$V(x)= a x^2 + b g x^3 +\ldots=\frac{1}{g^2}\,\hat{V}(gx)$, as well as for the radial oscillator $V(r)=\frac{1}{g^2}\,\hat{V}(gr)$ and for the perturbed Coulomb problem $V(r)=\frac{\al}{r}+ \beta g r + \ldots = g\,\tilde{V}(gr)$, the Perturbation Theory in powers of the coupling constant $g$ (weak coupling regime) and the semiclassical expansion in powers of $\hbar^{1/2}$ for the energies coincide. This is related to the fact that the dynamics developed in two spaces: $x\ (r)$-space and $gx\ (gr)$-space, lead to the same energy spectra. The equations which govern dynamics in these two spaces, the Riccati-Bloch equation and the Generalized Bloch equation, respectively, are presented. It is shown that the perturbation theory for the logarithmic derivative of the wavefunction in $gx\ (gr)$- space leads to (true) semiclassical expansion in powers of $\hbar^{1/2}$; for the one-dimensional case this corresponds to the flucton calculus for the density matrix in the path integral formalism in Euclidean (imaginary) time proposed by one of the authors \cite{Shuryak:1988}.
Matching the perturbation theory in powers of $g$ and the semiclassical expansion in powers of $\hbar^{1/2}$ for the wavefunction leads to a highly accurate local approximation in the entire coordinate space, its expectation value for the Hamiltonian provides a prescription for the summation of the perturbative (trans)-series.
\end{abstract}

\maketitle

\section*{Introduction}

The problem of bound states in non-relativistic quantum mechanics in ${\bf R^d}$ is governed by the time-independent Schr\"odinger equation
\begin{equation}
\label{SE}
  {\cal H}\ =\ -\frac{\hbar^2}{2 m} \sum_{i=1}^d \pa_{x_i}^2 + V(x)\quad ,\quad {\cal H}\Psi=E\Psi\quad ,\quad \int |\Psi|^2 d^d x < \infty\ ,
\end{equation}
where $V$ is the potential. In this paper we deal with two important classes of potentials:

\noindent
(I) the general one-dimensional anharmonic oscillator (AHO) potential,
\[
V_a(x)\ =\ \frac{1}{g^2}\,\hat{V}(gx)\ =\
\]
\begin{equation}
\label{AHO}
a_0^2 x^2 +
a_1 g x^3 + a_2 g^2 x^4 + \ldots +\ a_{k-2} g^{k-2} x^k + \ldots + a_{2p-2} g^{2p-2} x^{2p} + \ldots\ ,\quad x\in (-\infty, +\infty)\ ,
\end{equation}
where $V_a(x) \geq 0$; the quartic, sextic AHO as well as the quartic $x^4$, sextic $x^6$ etc oscillators, double-well, triple-well, quartic tilted (asymmetric) double-well potentials and even the sine-Gordon potential belong to this class among many other well-known potentials - as well as the $O(d)$,-symmetric radial anharmonic oscillator $$V_a(r)\ =\ \frac{1}{g^2}\,\hat{V}(gr)\ ,$$ and

\noindent
(II) the three-dimensional $O(3)$ radially-Perturbed Coulomb Problem (PCP)
\begin{equation}
\label{Coulomb}
V_c(r)\ =\ g\,\tilde{V}(gr) = -\frac{b_0}{r} + b_1 g^2 r + b_2 g^3 r^2 + \ldots \ ,\quad r \in [0,\infty)\ ,
\end{equation}
where the linear $(b_i=0, i>1)$ and quadratic $(b_i=0, i>2)$  funnel-type potentials as well as the celebrated Yukawa potential are among the potentials of this class. Without loss of generality in many occasions we can put $a_0=b_0=1$, also implying that the constant terms are absent in (\ref{AHO}), (\ref{Coulomb}) as a result of a specific choice of reference point for energy. In general, $\{a\}$ and $\{b\}$ are sets of real parameters. Needless to mention there is an enormous body of papers published about numerous particular cases of (\ref{AHO}) and (\ref{Coulomb}). In this paper we will focus on the one-dimensional anharmonic oscillators (\ref{AHO}) with polynomial anharmonicity of finite integer order $p$ and their corresponding ground state.

\section{Riccati-Bloch equation and Perturbation theory}

Consider the one-dimensional AHO as the first problem. Take the exponential representation of the wave function
\begin{equation}
\label{phase}
        \Psi\ =\ e^{-\frac{1}{\hbar}\phi} \ ,
\end{equation}
and substitute it into the Schr\"odinger equation (\ref{SE}) putting for simplicity $m=1/2$. We arrive at the well-known Riccati equation
\begin{equation}
\label{Riccati}
  \hbar\,y' \ -\  y^2\ =\ E\ -\  \frac{1}{g^2}\, {\hat V}(g x) \ , \quad y\,=\,\phi'\ ,
\end{equation}
see e.g. \cite{LL:1977}, which contains the Planck constant $\hbar$ in front of the leading derivative. Note that by ignoring this term (in the zeroth approximation), one arrives at the textbook WKB expression for $y$, see e.g. \cite{LL:1977}.


There are two ways to get rid of the explicit $\hbar$ dependence in this equation, which lead to two different expansions of its solution. One corresponds to the
{\it Riccati-Bloch equation}, created by making the following changes to the Riccati equation (\ref{Riccati}) - introducing a new variable $v$ and a new function $\mathcal{Y}$. We write it as
\begin{equation}
\label{RB-variables}
  x\ =\ {\hbar^{1/2}}\, v \ ,\ y\ =\ \hbar^{1/2}\, \mathcal{Y}\left(v\right)\ ,\ E\ =\ {\hbar}\,\veps\ ,
\end{equation}
and the effective coupling
\begin{equation}
\label{coupling}
   \la \ =\ \hbar^{1/2}\,g\ .
\end{equation}
In this case we arrive at the so-called {\it Riccati-Bloch (RB)} equation
\begin{equation}
\label{RB}
  \pa_v {\cal Y}\ -\ {\cal Y}^2\ =\ \veps (\la)\ -\ \frac{1}{\la^2}\,{\hat V}(\la v)\ , \quad \pa_v \equiv \frac{d}{dv}\ ,\ v\in (-\infty,\infty)\ ,
\end{equation}
where the potential remains of the same form as in (\ref{AHO}) with replacement $g \rar \la$,
\[
\frac{1}{\la^2}\,\hat{V}(\la v)\ =\ a_0^2\, v^2\ +\
a_1\, \la \,v^3\ +\ a_2\, \la^2\, v^4\ +\ \ldots\ +\ a_p\, \la^{2p-2}\, v^{2p}\ +\ \ldots \ .
\]
Formally, this equation has no $\hbar$-dependence: {\it we study dynamics in the ``quantum", $\hbar$-dependent coordinate $v$ (\ref{RB-variables}) instead $x$, which is governed by an $\hbar$-dependent, effective coupling constant $\la$ (\ref{coupling}).} If we develop the perturbation theory (PT) in powers of $\la$ in (\ref{RB}), putting $a_0=1$, for the ground state,
\begin{equation}
\label{PT-veps}
   \veps\ =\ \sum_0^{\infty} \la^{n} \veps_n\ ,\quad \ \veps_0\ =\ 1\ ,\ \veps_1\ =\ 0\ ,\ \ldots
\end{equation}
\begin{equation}
\label{PT-Y}
   {\cal Y}\ =\ \sum_0^{\infty} \la^{n} {\cal Y}_n(v)\ ,\quad \ {\cal Y}_0\ =\ v\ ,\ {\cal Y}_1\ =\
   \frac{a_1}{2}(v^2+1)\ ,\ \ldots
\end{equation}
it becomes clear that the expansion for the energy (\ref{PT-veps}) is simultaneously the perturbation series in powers of $g$ and the semiclassical expansion in powers of $\hbar^{1/2}$, since the coefficients $\veps_n$ are numbers, which depend on the parameters $a_{0,1,2,\ldots}$ in (\ref{AHO}), see (\ref{coupling}). In general, this expansion is asymptotic, having a zero radius of convergence: there exists the problem of its summation. This remains true for the AHO (\ref{AHO}) with two (or several) global degenerate minima. However, in this case (\ref{PT-veps}) contains exponentially-small terms in $\la$ in addition to the Taylor expansion in powers of $\la$. Contrary to the energy (\ref{PT-veps}), the expansion (\ref{PT-Y}) of ${\cal Y}$ is the PT expansion in powers of $g$ only, since the corrections $Y_n(v)$ are $\hbar$-dependent. This is {\it not} a semiclassical expansion in powers of $\hbar$.

It is worth noting that one can develop in Eq.(\ref{RB}) the perturbation theory in powers of $v$,
\[
    {\cal Y}\ =
\]
\begin{equation}
\label{Y-expansion}
   \al\ +\ (\veps\,+\,\al^2)\, v\ +\ \al(\veps+\al^2)\,v^2\ +\  \frac{\veps^2+\al^2 (4\veps+3\al^2)-a_0^2}{3}\, v^3 \ +\ \ldots \ ,
\end{equation}
where the parameters $\al\equiv{\cal Y}(0),\veps$ can be found approximately only. For even potentials the ground state function is even, its logarithmic derivative ${\cal Y}$ is odd, ${\cal Y}(-v)=-{\cal Y}(v)$ hence, $\al=0$.
Expansion (\ref{Y-expansion}) mimics the perturbation theory in powers of $\la$ for ${\cal Y}$ (\ref{PT-Y}): the coefficients in (\ref{Y-expansion}) can be found in the form of the expansion in powers of $\la$. We have to emphasize that the expansion (\ref{Y-expansion}) has the form of a Taylor expansion in $v$ even for one-term potentials
\[
V=\la^{2p-2}\, v^{2p}\ ,\ p=2,3,\ldots \ ,
\]
where the perturbation theory in powers of $\la$ can {\it not} be developed.

\section{ Generalized Bloch equation and semiclassical series}

Another approach, central to this paper, is best formulated in a new variable $u$, new function $\mathcal{Z}$ and new energy,
\begin{equation}
\label{GB-variables}
u\ =\ {g\,x}\ =\ \la v\ ,\quad y\ =\ \frac{1}{g}\mathcal{Z}(u)\ ,\quad E\ =\ {\hbar}\,\veps\ ,
\end{equation}
keeping the same effective coupling constant (\ref{coupling})
\[
   \la \ =\ \hbar^{1/2}\,g\ .
\]
After substitution of (\ref{GB-variables}) into the Riccati equation (\ref{Riccati}) and assuming $g \neq 0$ we arrive at
\begin{equation}
\label{GB}
   \la^2\,\pa_u\mathcal{Z}(u)\ -\ \mathcal{Z}^2(u)\ =\ \la^2\,\veps(\la)\ -\ {\hat V}(u)\quad ,\quad \pa_u\equiv\frac{d}{du}\ ,\ u \in (-\infty,\infty)\ ,
\end{equation}
cf.(\ref{RB-variables}), where
\[
  \hat{V}(u)\ =\ a_0^2\, u^2\ +\
a_1 \,u^3\ +\ a_2\, u^4\ +\ \ldots\ +\ a_p\, u^{2p}\ +\ \ldots \quad .
\]
This is the so-called {\it Generalized Bloch (GB)} equation, see \cite{ST:2018} for the case of the double well potential; evidently, it requires a regularization at $\la \rar 0$, when ${\hat V}(u)\rar u^2$, which eventually will lead to the RB equation. This equation describes dynamics in {\it classical, $\hbar$-independent} coordinate $u=g\,x$.

Now we develop the PT in powers of $\la$ in the equation (\ref{GB}) assuming that the AHO potential (2) has no degenerate global minima. It is evident that the expansion of the energy $\veps$ in powers of $\la$ (\ref{PT-veps}) remains the same as in the Riccati-Bloch equation (\ref{RB}), unlike the expansion for ${\mathcal{Z}}$, which becomes different,
\begin{equation}
\label{PT-Z}
   {\cal Z}\ =\ \sum_n \la^n {\cal Z}_n(u)\ ,\ {\cal Z}_0\ =\ {\sqrt{ {\hat V} (u)}}\ ,\ {\cal Z}_1\ =\ 0\ ,
   \ {\cal Z}_2\ =\ \frac{1}{2} \left(\log {\sqrt{ {\hat V} (u)}}\right)^{\prime}_u - \frac{\veps_0}{2{\sqrt{ {\hat V} (u)}}}\ ,\ \ldots \ .
\end{equation}
where ${\veps_0}=1$. Note that in the standard WKB approach ${\veps_0}$ is replaced by $\veps$, which depends on $g$ and $\hbar$, and, in general, it can {\it not} be found exactly. It is worth presenting three particular examples:

(i) the quartic anharmonic oscillator ${\hat V}=u^2+u^4$, where
\[
  {\cal Z}_0\ =\ {u \sqrt{1+u^2}}\ ,\ {\cal Z}_2\ =\ \frac{1}{4} \left(\log u^2 (1+u^2) \right)^{\prime}_u - \frac{1}{2{u \sqrt{1+u^2}}}\ ,
\]

(ii) the sine-Gordon potential ${\hat V}=\sin^2 u$, where
\[
  {\cal Z}_0\ =\ \sin u \ ,\ {\cal Z}_2\ =\ \frac{1}{2} \cot u - \frac{1}{2\,{\sin u}}\ ,
\]
and,

(iii) the quartic oscillator ${\hat V}=u^4$, where
\[
  {\cal Z}_0\ =\ {u|u|}\ ,\ {\cal Z}_2\ =\ \frac{1}{u} - \frac{\veps_0}{2{u|u|}} \ ,
\]
here $\veps_0 \approx 1.0604$ has the meaning of the ground state energy of the quartic oscillator.

It can be immediately recognized that ${\cal Z}_0$ is, in fact, the classical momentum {\it at zero energy} when $\la=1$. In turn, $\int {\cal Z}_0\ du$ is the classical action at zero energy. Furthermore, replacing in ${\cal Z}_0$ the argument $u=\la v$ (\ref{GB-variables}) one can see that ${\cal Z}_0(\la v)$ is the generating function for the leading terms of the highest degrees in $v$ of the ${\cal Y}_n(v)$ corrections of the expansion (\ref{PT-Y}), while ${\cal Z}_2(\la v)$ is the generating function for the next, subleading terms of the ${\cal Y}_n(v)$ corrections etc. Note that
\begin{equation}
\label{eqn_det}
   \int {\cal Z}_2\ du\ =\ \frac{1}{4}\,{\log { {\hat V} (u)}}\ -\
   \frac{{\veps_0}}{2} \int \frac{du}{\sqrt{{\hat V}(u)}}\ ,
\end{equation}
is related to the logarithm of the determinant, see Section IV. In the standard WKB expansion
the first term in the rhs of (\ref{eqn_det}) contains the energy while the second term is absent.

In general, expansion (\ref{PT-Z}) is the true semiclassical expansion in powers of  $\hbar^{1/2}$,
\[
     {\cal Z}\ =\ \sum_n  \hbar^{\frac{n}{2}} \big( g^n\,{\cal Z}_n(g x) \big)\ ,
\]
while
\[
      \veps(\hbar^{1/2}\,g)\ =\ \sum^{\infty}_{n=0} (\hbar^{1/2}\,g)^{n} \veps_n\ .
\]

It is a well-known fact that if the potential (\ref{AHO}) has two or more degenerate global minima, e.g. $V_{dw}=x^2(1-gx)^2$, in addition to the Taylor expansion in powers of $\la$: $E_{PT}(\la)=\sum E_n \la^n$, the exponentially-small terms  occur, which can be summed into the non-perturbative energy $E_{NPT}$. In particular, for the case of double-well potential $V_{dw}$, the energy of the state can be written as $E=E_{PT}+E_{NPT}$ \cite{ST:2018}. Hence, it manifests the occurrence of trans-series in $\la$, see e.g. \cite{Shifman:2015}, and also \cite{ST:2018} (and references therein), instead of the Taylor expansion.

The RB equation (\ref{RB}) continues to hold and one can see immediately that the energy depends on the single combination of parameters: $\la \ =\ \hbar^{1/2}\,g$,
\[
     \veps (\hbar^{1/2}\,g)\ =\ \veps_{PT} (\hbar^{1/2}\,g) + \veps_{NPT} (\hbar^{1/2}\,g)\ .
\]
Hence, the semiclassical expansion in powers of $\hbar^{1/2}$ (\ref{PT-veps}) becomes the semiclassical expansion in the form of a trans-series in $\hbar^{1/2}$.

It is worth noting that for a polynomial potential, where the expansion (\ref{AHO}) is terminated at degree $(2p)$, its expansion in $u$-variable takes the form
\begin{equation}
\label{AHO-2p-u}
  \hat{V}(u)\ =\ a_0^2\, u^2\ +\
a_1 \,u^3\ +\ a_2\, u^4\ +\ \ldots\ +\ a_{k-2} u^k\ +\ \ldots\ +\ a^2_{2p-2}\, u^{2p} \quad .
\end{equation}
In Eq.(\ref{GB}) one can develop the asymptotic expansion in inverse powers of $u$,
\begin{equation}
\label{Z-expansion}
    {\cal Z}\ =\ \pm\, a_{2p-2}\, u^p\ +\ \frac{a_{2p-3}}{\pm \,2\, a_{2p-2}}\, u^{p-1}\ +\ \frac{1}{2} \bigg(\frac{a_{2p-4}}{a_{2p-3}}+\frac{a_{2p-3}}{4a_{2p-2}^2}\bigg)\,u^{p-2}\ +\ \ldots\ +\ \frac{\la^2p+ \frac{a_{p-3}}{\pm a_{2p-3}}}{2u}\ +\ \ldots \quad ,
\end{equation}
where for even $p$ the sign {\it plus} is chosen for positive $u > 0$ and the sign {\it minus} for negative $u < 0$ to assure square integrability of the ground state function. 
If $p$ is odd, the sign is always plus.
It is evident that the first $(p)$ coefficients in expansion (\ref{Z-expansion}) do not depend on $\la$, while the first $(2p)$ coefficients do not depend on the energy. Hence, the first $(2p)$ coefficients in expansion (\ref{Z-expansion}) can be found exactly. This expansion mimics the perturbation theory expansion in powers of $\la$ for ${\cal Z}$ (\ref{PT-Z}), when the corrections ${\cal Z}_n$ are expanded in $1/u$.

Interesting situation occurs when AHO degenerates to a power-like potential, thus, all $a_0=a_1=\ldots=a_{2p-3}=0$ except for $a_{2p-2}\neq 0$. In this case the potential (\ref{AHO-2p-u}) has the form
\[
    V \ =\ a^2\,u^{2p} \ ,\ a^2 \equiv a_{2p-2}\ .
\]
In expansion (\ref{Z-expansion}) all terms of non-negative degrees vanish except for the leading degree $p$ as well as all negative degrees $-2,-3,\ldots -(p-1), -(p+1)$\ ,
\begin{equation}
\label{Z-expansion-1term}
    {\cal Z}\ =\ \pm\, a\, u^p\ +\ \frac{\la^2 p}{2}\ \frac{1}{u}\ -\ \frac{\la^2 \veps}{\pm 2 a} \frac{1}{u^p}\ - \ \frac{\la^2 p(p+2)}{\pm 8 a}\ \frac{1}{u^{p+2}} \ + \
    \ldots \quad ,
\end{equation}
where for the case of even $p$ the sign plus (upper sign) is chosen for positive $u > 0$ and the sign minus (lower sign) for negative $u < 0$ to assure square integrability of the ground state function. For the case of odd $p$ the sign in (\ref{Z-expansion-1term}) should be always chosen plus.

The property, that the first $(2p)$ coefficients in expansion (\ref{Z-expansion}) (and (\ref{Z-expansion-1term})) do not depend on the energy and can be found exactly in terms of the parameters of the potential, plays crucially important role in the construction of the approximation for ground state wavefunction, see Section \ref{matching}.

\section{Semiclassical approximation of path integrals, the ``flucton" paths}

Following Feynman \cite{F-H}, the amplitude of a quantum system to go from point $x_i$ to point $x_f$ in time $t$ can be expressed as a functional integral
over all paths, starting and ending at these points.
He also famously pointed out that by moving this expression to Euclidean (imaginary) time $\tau=it$ defined on a circle with ``Matsubara time" circumference related to temperature \begin{equation}
    \beta\ =\ \frac{\hbar}{T}\ ,
\end{equation}
one can use path integrals in statistical mechanics. Specifically, the partition function is given by integrals over the periodic paths with time period $\beta$.

In this paper we will use this formalism in the zero temperature limit only, in which $\beta \rightarrow \infty$ and the path integral naturally describes the ground state. Its density matrix -- the probability to find quantum particle at certain position $P(x_0)$ --
is given by the integral over periodic paths, where the initial and final points coincide, $x_i=x_f=x_0$. In the $\beta \rightarrow \infty$ limit it becomes the square of the ground state wave function $P(x_0)=|\psi_0(x_0)|^2$. Thus,
\begin{equation}
\label{psi0-2}
  \psi_0^2(x_0)\ =\ \int {\cal D}x\ e^{-\frac{S_E}{\hbar}}\ ,
\end{equation}
where $S_E$ is the Euclidian action, if $\psi_0(x)$ is a real, positive function. Needless to say that the square-root of the rhs is the exact solution of the Schr\"odinger equation.

The development of the semiclassical theory of the ground state, based on special classical paths called ``fluctons",  started from the  paper of one of the present authors \cite{Shuryak:1988}.
Significant development of this theory has been then made in two recent papers \cite{Escobar-Ruiz:2016,Escobar-Ruiz:2017}, focused on the ground state of a number of quantum-mechanical problems, harmonic and anharmonic oscillators, as well as the double-well
and sine-Gordon  potentials \footnote{
For first application of this semiclassical approximation to finite temperatures for
some of these examples, see \cite{Shuryak:2019}.}.

A {\it Flucton} is a path possessing the least action among all paths passing via the {\em observation point} $x_0$. Therefore it satisfies the classical (Newtonian) equation of motion
\begin{equation}
\label{NE}
     m \ddot x(\tau)={\pa V \over \pa x}\ ,
\end{equation}
where the dots indicate derivatives over $Euclidean$ time $\tau$. Note that therefore the usual minus sign in the r.h.s. is absent: one can view this as motion in the {\it inverted} potential $\left(-V(x)\right)$.

As usual in $1D$ mechanical problems, in order to find trajectory in (\ref{NE}) the easy way is to employ the energy conservation, which in this case takes the form,
\begin{equation}
 \frac{m}{2}\,\dot x(\tau)^2 -V(x)\ =\ E\ .
\end{equation}
The maximum of the inverted potential $(-V(x))$ is conveniently put to zero, so a particle with $E=0$ may stay at this maximum for infinite time. This simple idea defines the shape of the {\it flucton}.

Before giving its explicit form, let us do some redefinitions. For finite $\beta$, the
time variable $\tau$ is defined on a circle. The only condition on paths is
that they must pass through the observation point $x_0$, but it does not matter
at what moment in time this happens. Therefore, one can define it to be zero
\begin{equation}\label{eqn_condition}
x(\tau=0)=x_0\ ,
\end{equation}
with two symmetric arms, for positive and negative $\tau$,
describing {\it path relaxation}. In the zero temperature, or $\beta\rightarrow \infty$,
limit we discuss,  $\tau \in (-\infty,\infty)$ and the asymptotic coordinate values
correspond to the position of the potential minimum, defined as
\[
x(\tau\rightarrow \pm \infty) =0\ .
\]
The explicit functional shape follows readily from energy conservation (with $m=1, E=0$)
\begin{equation}
\label{eqn_tau}
  \tau=\int_{x_0}^{x(\tau)} \frac{dx}{\sqrt{2V(x)} }\ .
\end{equation}

At this point let us break presentation and, following the wisdom of new variables introduced in the preceding section, describe the Euclidean action. With the classical coordinate $u=g\,x$ and the potential $V(x)=\hat V(u)/g^2$, it takes the form
\begin{equation}
\label{action}
     \frac{S}{\hbar}\ =\ \frac{1}{\hbar g^2}\,\int d\tau
\bigg(\frac{1}{2}\dot u(\tau)^2 +\hat V(u)\bigg) \ ,
\end{equation}
see (\ref{psi0-2}),
in which the coupling is united with the Planck constant (leading to the effective coupling $\la^2=\hbar g^2$), but both of them are absent in the Equation of Motion (EoM).
The path integral now takes the form
\begin{equation}
\label{path-int}
P(u_0)    =\ \int {\cal D} u\ e^{-\frac{1}{\hbar g^2}\,\int d\tau
\bigg(\frac{1}{2}\dot u(\tau)^2 +\hat V(u)\bigg)} \ ,
\end{equation}
where, we remind, the dependence on the observation point $u_0=g x_0$
comes from the condition (\ref{eqn_condition}) which paths must obey.

The change of variable in (\ref{action}) to the ``classical variable" $u$ has important consequences.
In particular,  the integral in the r.h.s. of (\ref{eqn_tau}) becomes
\begin{equation}
\label{eqn_tau-u}
      \tau\ =\ \int_{u_0}^{u(\tau)} \ \frac{du}{\sqrt{2 \hat V(u)} }\ ,
\end{equation}
independent of $g$ and $\hbar$, where $u_0=g x_0$ is the starting point in $u$-space.
Therefore, in these notations the flucton shape is universal, it does not depend on the coupling constant.

\begin{figure}
\centering
    \includegraphics[width=10cm]{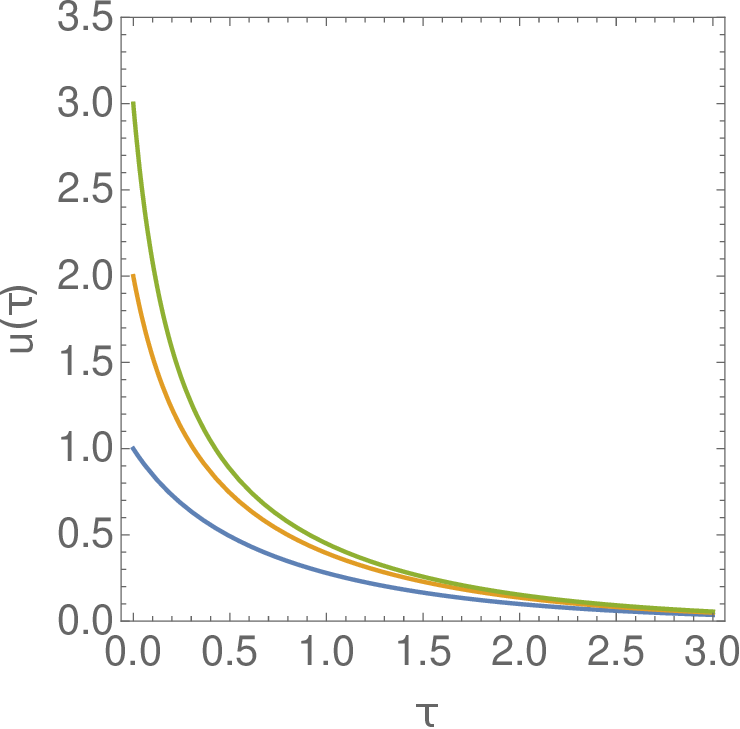}
 	\caption{Examples of different flucton paths $u_{fl}(\tau)$ (\ref{ufl-24}),
    when $u(0)=1,2,3$ versus Euclidean time $\tau$ (blue, yellow, green curves, respectively). Only half of the path, for positive $\tau$ is shown: the path at negative values is its mirror image. 	}	
 	\label{fig:fluctons}
\end{figure}

For an example at hand, $\hat V=u^2/2+u^4$, the integral (\ref{eqn_tau-u}) can be evaluated analytically, its inverse can also be found in closed analytic form, giving
\begin{equation}
\label{ufl-24}
   u_{fl}(\tau)\ =\ \frac{1}{\sqrt{2} sinh\,\big| arc csch\big(\sqrt{2}u_0 \big) +\tau
   \big| }\ .
\end{equation}
Three examples of flucton paths for different $u(0)$ are plotted in Fig.\ref{fig:fluctons}.
Inserting (\ref{ufl-24}) into the (Euclidean) action (\ref{action}), one gets
\begin{equation}
\label{Sfl-ufl-24}
    S_{fl}\ = \ \frac{1}{6\,{\hbar}\,g^2}\, \bigg(-1 + {(1 + 2 u_0^2)}^{3/2} \bigg)  \ ,
\end{equation}
and therefore the explicit form of the density matrix $P(u_0)\sim exp\big(-S_{fl})$.
This result, of course, reproduces the standard WKB expression for the ground state wave function at zero energy $E=0$, cf.(\ref{PT-Z}).

\begin{figure}
\centering
    \includegraphics[width=10cm]{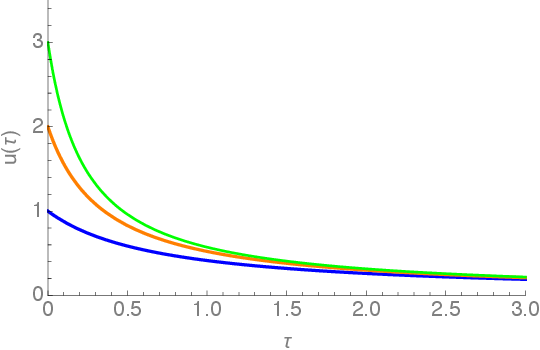}
 	\caption{Examples of flucton paths $u_{fl}(\tau)$ (\ref{ufl-4}),
    when $u(0)=1,2,3$ versus Euclidean time $\tau$ (blue, yellow, green curves, respectively). Only half of the path, for positive $\tau$ is shown: the path at 
    negative values is its mirror image. 	}	
 	\label{fig:fluctons-4}
\end{figure}

As for another example, for instance, the quartic oscillator, ${\hat V}=u^4$, the integral in the r.h.s. of (\ref{eqn_tau-u}) can also be found in closed analytic form, leading to
\begin{equation}
\label{ufl-4}
       u_{fl}=\frac{u_0}{1+{\sqrt 2} u_0 \tau} \ ,
\end{equation}
which is not much different than the flucton trajectory for the anharmonic oscillator,
cf.(\ref{ufl-24}). Three examples of flucton paths for the quartic oscillator for different $u(0)$ are plotted in Fig.\ref{fig:fluctons-4}. Putting (\ref{ufl-4}) into the (Euclidean) action (\ref{action}), one gets
\begin{equation}
\label{Sfl-ufl-4}
    S_{fl}\ = \ \frac{\sqrt{2}}{3\,{\hbar}\,g^2}\,u_0^3  \ ,
\end{equation}
cf.(\ref{Sfl-ufl-24}), and therefore the explicit form of density matrix $P(u_0)\sim exp\big(-S_{fl})$.

However, unlike standard WKB, the flucton version of the semiclassical theory allows one to derive systematically the series of the semiclassical corrections in powers of $\hbar^{1/2}$, see \cite{Escobar-Ruiz:2016,Escobar-Ruiz:2017}. This works as follows. Arbitrary path can be viewed as a classical flucton path plus a quantum fluctuation
\begin{equation}
   u(\tau)=u_{fl}(\tau)+ \hbar^{1/2} q(\tau) \ ,
\end{equation}
By substituting this into the action one can do systematic expansion in powers of $q$
\begin{equation}
\label{action-SS}
 \frac{S-S_{fl}}{\hbar}\ =\ \frac{1}{{ \hbar\,g^2}} \int d\tau \bigg(\frac{{\dot q}^2}{2}\ +\
 \frac{1}{2!}\,\frac{\pa^2 \hat V(u=u_{fl})}{\pa u^2}\, q^2(u)\ +\
 \frac{\hbar^{1/2}}{3!}\,\frac{\pa^3 \hat V(u=u_{fl})}{\pa u^3}\, q^3(u)\ +\ \dots \bigg)\ .
\end{equation}
Note that for the first two terms (quadratic in $q$) both $g$ and $\hbar$ are absent, while the subsequent nonlinear terms contain growing powers of $\hbar^{1/2}$.
If only the quadratic terms in $q$ are kept, the resulting EoM is linear, defining the so called ``fluctuation operator" $\hat O$. In this approximation the functional integral over the fluctuations is Gaussian, producing the {\it determinant} of $\hat O$ operator. In the notations we use it becomes obvious that the operator, its eigenvalue spectrum, and their product -- the determinant -- are all universal,  independent of $g$ and $\hbar$.

It is important, that unlike instantons and many other classical trajectories (e.g. solitons), the flucton background does not have $any$ zero modes of $\hat O$, and  therefore its inversion is straightforward. Indeed, there is $no$ symmetry corresponding to the time shift, because  by definition the paths should satisfy condition (\ref{eqn_condition}), which
implies that $q(0)=0$.

The eigenvalue equation
\[
     \hat O\, u_\la(\tau)\ =\ \la\, u_\la(\tau) \ ,
\]
is of second order in the derivative and thus similar to the Schr\"odinger equation with an effective potential given by the second derivative at the background. For the example at hand, $\hat V=u^2/2+u^4$ (the quartic anharmonic oscillator),
\[
     \hat V''(u=u_{fl})\ =\ 1+12 u_{fl}^2\ ,
\]
where $u_{fl}$ is presented in (\ref{ufl-24}),
hence, the effective potential is equal to 1 at $\tau \rightarrow \pm \infty$ and larger than 1 at the origin, $\tau=0$. Clearly, there are no bound states and all states are scattering states. With standard quantization in a box, all those can be found. This direct diagonalization approach has been used in our previous works \cite{Escobar-Ruiz:2016,Escobar-Ruiz:2017}.
However, this is no longer necessary, as in Section II we have derived the analytic expression for the determinant for any arbitrary potential(!), see (\ref{eqn_det}).

The higher order terms $O(q^3,q^4...)$ in the expansion (\ref{action-SS}) can be viewed as vertices of Feynman diagrams: the growing powers of $\hbar^{1/2}$ in front of them show that it is a truly semiclassical expansion. The propagators which occur in such diagrams are nothing but the Green function, which is the inverse of the fluctuation operator $\hat O$. In fact, several of them were calculated in Section II, see (\ref{PT-Z}). These terms correspond to 1-,2-,3-... loop results. It is worth noting that the one-loop contribution is given (formally) by a single diagram, see \cite{Escobar-Ruiz:2016}, Fig.3 and \cite{Escobar-Ruiz:2017}, Fig.2, the two-loop contribution is the sum of three diagrams, see \cite{Escobar-Ruiz:2016}, Fig.4 and \cite{Escobar-Ruiz:2017}, Fig.3, etc. A non-trivial fact observed in calculations of concrete systems \cite{Escobar-Ruiz:2016,Escobar-Ruiz:2017} is that each individual diagram can be quite complicated, can contain highly transcendental expressions, not always can be calculated analytically, while the sum of the individual diagrams of a given order leads to some mysterious cancellations which results in a sufficiently simple final expressions. A certain advantage of the formalism developed in Section II is that the expansion (\ref{PT-Z}) deals with sum of Feynman diagrams directly 
and we do {\it not} see these complications.

\section{Matching perturbation theory and semiclassical expansion}
\label{matching}

Let us consider the polynomial potential of degree $(2p)$, see (\ref{AHO-2p-u}).
By taking the perturbation theory for the logarithmic derivative ${\cal Y}$ (\ref{PT-Y}) at small distances (or, saying differently, the asymptotic expansion (\ref{Y-expansion})) and matching it with a new version of the semiclassical expansion of ${\cal Z}$ (\ref{PT-Z}) at large distances (or, saying differently, the asymptotic expansion (\ref{Z-expansion})) we arrive practically unambiguously(!) at approximate eigenfunction for the $k$th excited state of the form

\begin{equation}
\label{approximant}
   \Psi^{(k)}(x)\ =\ P_k(x)\ \mbox{prefactor} (x,g;\{ {\tilde b}\})
   \exp{ \bigg(\ -\ \frac{{A}\ +\
   {\hat V}(x,g; \{ {\tilde a}\}) }{\sqrt{\frac{1}{x^{2}}\,{\hat V}(x,g;\ \{ {\tilde b}\} )}}
   \ +\ \frac{A}{{\tilde b}_0^2}\bigg)} \ .
\end{equation}
Here ${\hat V(x,g;\ \{ {\tilde a}\} )}$, ${\hat V(x,g;\ \{ {\tilde b}\})}$ are, as defined in (\ref{AHO}), the potentials but with new and different sets of coefficients $\{ {\tilde a}\}$ or $\{ {\tilde b}\}$, respectively. In general, $\{ {\tilde a}\},  \{ {\tilde b}\}$ and $A$ are free (variational) parameters subject to $(p)$ constraints: for the phase $\phi$ (\ref{phase}) all $(p)$ growing terms at large distances do not depend on energy and should be reproduced exactly. Here $P_k(x)$ is a polynomial of degree $k$, all its coefficients are found by imposing the orthogonality condition for the functions $\Psi^{(\ell)}(x)$, see (\ref{approximant}), with $\ell=0,1,2,\ldots, (k-1)$. The prefactor in (\ref{approximant}) depends on the potential, it is usually defined by ${\cal Z}_2$ in (\ref{PT-Z}) (by the determinant), see also (\ref{eqn_det}).

Formula (\ref{approximant}) is the central formula of this article. Let us present two examples. It is easy to check that for the Harmonic Oscillator (HO),
\[
        V_{HO}\ =\ a_0^2 x^2\ ,
\]
cf.(\ref{AHO}), the (\ref{approximant}) becomes exact, with no free parameters,
\[
        \Psi_{HO}\ =\ P_k\,e^{-\frac{a_0}{2} x^2}\ ,
\]
where the prefactor is absent and $P_k$ is the $k$th Hermite polynomial. Another example is
the quartic symmetric AnHarmonic Oscillator (AHO),
\[
        V^{(4)}_{AHO}\ =\ a_0^2 x^2 + a_2^2 g^2 x^4\ ,
\]
where we can set $a_0=a_2=1$.
As a result of matching the two asymptotic expansions at small and large distances (equivalently, the perturbation theory in $x$ and the new semiclassical expansion)
we arrive at the following function for the $(k=2n+p)$-excited state with quantum numbers $(n,p)$, $n=0,1,2,\ldots\ ,\ p=0,1$\ \footnote{in this case $k$ is the {\it principal} quantum number} as a reduction of (\ref{approximant}) with two  imposed constraints, emerging from (\ref{Z-expansion}) \cite{Turbiner:2021},
\[
 \Psi^{(n,p)}_{(approximation)}\ =\
 \frac{x^p P_{n,p}(x^2; g^2)}{\left(B^2\ +\ g^2\,x^2 \right)^{\frac{1}{4}}
 \left({B}\ +\ \sqrt{B^2\ +\ g^2\,x^2} \right)^{2n+p+\frac{1}{2}}}
\]
\begin{equation}
\label{final}
   \exp \left(-\ \dfrac{A\ +\ (B^2 + 3)\,x^2/6\ +\ g^2\,x^4/3}
  {\sqrt{B^2\ +\ g^2\,x^2}} \ +\ \frac{A}{B}\right)\ ,
\end{equation}
where $P_{n,p}$ is some polynomial of degree $n$ in $x^2$ with positive roots. Here $A=A_{n,p}(g^2),\ B=B_{n,p}(g^2)$ are two variational parameters. It was shown in \cite{Turbiner:2021} that for the six lowest states $n=0,1,2$ and $p=0,1$ the variational energies for coupling constants in $g \in [0,\infty)$ are obtained with an accuracy of 10-11 significant digits. The variational (optimal) parameters $A=A_{n,p}(g^2),\ B=B_{n,p}(g^2)$ are easily fitted \cite{Turbiner:2021} leading to an accuracy in energy (the expectation value for the Hamiltonian) of 9-10 significant digits for any coupling constant $g \geq 0$.


\vskip .8cm

{\it \bf CONCLUSIONS}

\vskip 1cm

It is shown that for the family of perturbed harmonic oscillators $V_a$ (\ref{AHO}), which includes the sine-Gordon potential, in particular, the classical coordinate $u=g x$ can naturally be introduced. This leads to the fact that the effective coupling constant of the theory becomes a combination of the Planck constant $\hbar$ and the coupling constant $g$: $\hbar^{1/2} g$. As a result the action is independent on external parameters: it contains the effective coupling constant as a factor in front of it only. It is evident that a similar phenomenon occurs for the radially perturbed radial harmonic oscillator, see \cite{delValle} and the radially perturbed Coulomb problem $V_c$ (\ref{Coulomb}), this will be presented elsewhere. It is worth noting that an analogue of the classical coordinate - the {\it classical field} - exists for the massless scalar field theory $\lambda \phi^{2n}$, in QED and in the Yang-Mills theory.

In quantum mechanics the Schr\"odinger equation for the potential $V_a$ (\ref{AHO}) can be transformed into the so-called generalized Bloch equation for the logarithm of the wave function. In this equation the perturbation theory in powers of $g$ coincides with the semiclassical expansion in powers of $\hbar^{1/2}$.
This allows us to easily construct the loop expansion of the density matrix in the path integral formalism dealing with sums of Feynman integrals with a given number of loops, in particular, to calculate the determinant for the general potential $V_a$ (which is the one loop contribution) in closed analytic form. The existence of the field theoretical analogue of the generalized Bloch equation remains an open question. The only available option (at the present moment) for developing the semiclassical expansion is to construct a loop expansion over the flucton background in field theory via calculation of the individual Feynman diagrams. For the case of the massless $\lambda \phi^{4}$ theory, using Lipaton \cite{Lipatov:1977} as the flucton trajectory, this will be done elsewhere.

\begin{acknowledgments}
A.V.T. gratefully acknowledges support from the Simons Center for Geometry and Physics,
Stony Brook University at which the research for this paper was initiated, it was supported
in part by DGAPA grant {\bf IN113022 }~(Mexico). The work of E.S. is supported in part by the U.S. Department of Energy, Office of Science under Contract No. {\bf DE-FG-88ER40388}.
A.V.T. thanks M A Shifman for useful discussions.

\end{acknowledgments}


\end{document}